\documentclass[floatfix,nofootinbib]{revtex4-2}

\usepackage{graphicx}
\usepackage{dcolumn}
\usepackage{bm}

\usepackage{color}
\usepackage{amsmath}
\usepackage{amsfonts}
\usepackage{amssymb}
\usepackage{latexsym}
\usepackage{caption}
\usepackage{subcaption}
\usepackage{tikz}
\usepackage{circuitikz}
\usepackage{hyperref}
\definecolor{darkred}{rgb}{0.8,0.1,0.1}
\hypersetup{colorlinks=true, linkcolor=darkred, citecolor=blue, linktoc=page}
\usepackage{tikz-cd,mathtools}

\def\m{\mu}
\def\n{\nu}

\def\){\right)}
\def\({\left( }
\def\]{\right] }
\def\[{\left[ }

\newcommand{\be}{\begin{equation}}
\newcommand{\ee}{\end{equation}}

\def\bsb{{\boldsymbol{b}}}
\def\bsk{{\boldsymbol{k}}}

\def\bsq{{\boldsymbol{q}}}

\def\a{\alpha}

\def\no{\nonumber}

\tikzset{graviton/.style={decorate, decoration={snake, amplitude=.4mm, segment length=1.5mm, pre length=.5mm, post length=.5mm}, double}}

\begin{document}

\title{
Gravitational wave double copy of radiation from gluon shockwave collisions
}

\author{Himanshu Raj}
\email{himanshu.raj@stonybrook.edu}
\affiliation{
Center for Frontiers in Nuclear Science, Department of Physics and Astronomy, Stony Brook University, Stony Brook, NY 11794, USA
}

\author{Raju Venugopalan}
 \email{raju.venugopalan@gmail.com}
\affiliation{
Department of Physics, Brookhaven National Laboratory, Upton, NY 11973, USA\\
Center for Frontiers in Nuclear Science, Department of Physics and Astronomy, Stony Brook University, Stony Brook, NY 11794, USA}

\date{\today}

\begin{abstract}
In \cite{Raj:2023irr}, we showed that the gravitational wave spectrum from trans-Planckian shockwave scattering in Einstein gravity is determined by the  gravitational Lipatov vertex expressed as the bilinear double copy $\Gamma^{\m\n} = \frac12 C^\m C^\n - \frac12 N^\m N^\n$ where $C^\m$ is the QCD Lipatov vertex and $N^\m$ is the QED soft photon factor. We show here that this result can be directly obtained by careful application of the classical color-kinematic duality to the spectrum obtained in gluon shockwave collisions. 
\end{abstract}

\maketitle

The trans-Planckian amplitude for the perturbative $2\rightarrow N$ scattering of two gravitons to $N\gg 1$ gravitons was computed in Regge asymptotics by Lipatov over forty years ago~\cite{Lipatov:1982it,Lipatov:1988ii}. In so-called multi-Regge kinematics, this amplitude can be decomposed into blocks of $2\rightarrow 3$ amplitudes determined by an effective ``Lipatov" vertex with the $t$-channel exchange of dressed (``reggeized") gravitons\footnote{Identical results were subsequently obtained in a 2-D effective theory of gravity~\cite{Amati:1987uf,Lipatov:1991nf,Amati:1993tb}.}. The computation is completely analogous to the framework for $2\rightarrow N$ gluon scattering in perturbative QCD that is described by the well-known BFKL equation~\cite{Kuraev:1977fs,Balitsky:1978ic}. Further, Lipatov showed that the effective $2\rightarrow 3$ gravitational vertex could be expressed as 

\begin{align}
\label{LipatovVertexGrav}
    \Gamma^{\m\n}(\bsq_1, \bsq_2) = \frac12 C^\m(\bsq_1, \bsq_2) C^\n(\bsq_1, \bsq_2) - \frac12 N^\m(\bsq_1, \bsq_2) N^\n(\bsq_1, \bsq_2)~.
\end{align}
Here $C^\m$ is the corresponding effective vertex for the $2\to 3$ gluon scattering amplitude in multi-Regge kinematics\footnote{A pedagogical  review of this formalism can be found in \cite{DelDuca:1995hf}.}. 
Its explicit form is given by
\begin{align}
\label{LipatovVertexQCD}
C^\m(\bsq_1, \bsq_2) \simeq -\bsq_{1}^\m+\bsq_{2}^\m + p_{1}^\m\left(\frac{p_2 \cdot k}{p_1 \cdot p_2}-\frac{\bsq_1^2}{p_1 \cdot k}\right)-p_{2}^\m\left(\frac{p_1 \cdot k}{p_1 \cdot p_2}-\frac{\bsq_2^2}{p_2 \cdot k}\right)~,
\end{align}
where $k$ is the four-momentum of the produced gluon and $\simeq$ indicates that only the transverse components of the momenta $q_1,q_2$ of exchanged gluons between incoming gluons with incoming four-momenta $p_1,p_2$ are relevant in  multi-Regge kinematics. The soft photon factor $N^\m$ in Eq.~\eqref{LipatovVertexGrav} is given by 
\begin{align}
N^\m (\bsq_1,\bsq_2)=\sqrt{\bsq_1^2 \bsq_2^2} \left(\frac{p_{1}^\m}{p_1\cdot k}-\frac{p_{2}^\m}{p_2\cdot k}\right)~.
\end{align}
It plays an important role in the gravitational scattering amplitude by removing unphysical simultaneous poles in the overlapping $s_1 = (k+p_1)^2$ and $s_2 = (k+p_2)^2$ channels of the $2\rightarrow 3$ sub-amplitude. 

In \cite{Raj:2023irr}, we showed that Eq.~\eqref{LipatovVertexGrav} is recovered in the computation of the inclusive gravitational wave spectrum in the trans-Planckian scattering of gravitational shockwaves. Our computation was analogous to Yang-Mills computations of gluon radiation in shockwave collisions where the spectrum is determined by $C^\mu$, the QCD Lipatov vertex~\cite{Blaizot:2004wu,Gelis:2005pt}. Such gluon shockwave collisions are realized in ultrarelativistic heavy-ion collisions and their dynamics leads to the formation of a quark-gluon plasma~\cite{Berges:2020fwq}.

In this note, we will show that Eq.~\eqref{LipatovVertexGrav} can be recovered directly by applying the classical double copy computations~\cite{Monteiro:2014cda,Goldberger:2016iau} relating Yang-Mills theory to Einstein gravity that employ a Bern-Carrasco-Johansson (BCJ)-type color-kinematic duality~\cite{Bern:2008qj,Bern:2019prr}. This classical double copy relation is due to Goldberger and Ridgway \cite{Goldberger:2016iau} who showed that the radiation produced in the scattering of colored scalar point particles, interacting via Yang-Mills fields, admits a double copy relation in the form of a color-kinematic replacement. This replacement, discussed later in Eq.~\eqref{DCprescription}, can be used to compute radiation generated by the scattering of scalars in dilaton gravity\footnote{A double copy of pure Yang-Mills alone, without colored scalar particles, gives Einstein gravity coupled to a dilaton and an antisymmetric Kalb-Ramond field. This is seen by counting the physical degrees of freedom. The Kalb-Ramond field does not appear in the current discussion because scalars do not couple to them unlike external particles with spin. Spinning particles do have  Lorentz and gauge invariant couplings with the antisymmetric field \cite{Goldberger:2017ogt}. While the double copy proposed in \cite{Goldberger:2016iau} relates pure YM to dilaton gravity, the dilaton decouples in ultrarelativistic limit. This limit will be the appropriate one for our discussion of the Lipatov vertex.}.

Our starting point is the leading order result for the classical solution of the Yang-Mills radiation field in the collision of two colored charges $c_\alpha^a(\tau_\alpha)$ of mass $m_\alpha$  and arbitrary velocities $v^\alpha$ ($\alpha=1,2$), where $a$ the color index and $\tau_\alpha$ parameterizes the respective worldline trajectories. The result for the radiation field is~\cite{Goldberger:2016iau}
\begin{equation}
\label{GRMain}
\begin{gathered}
A^{\mu,a}(k) =-\frac{g^3}{k^2} \sum_{\substack{\alpha, \beta = 1,2 \\
\alpha \neq \beta}} \int \mu_{\alpha, \beta}(k)\left[\frac{c_\alpha \cdot c_\beta}{m_\alpha} \frac{q_\alpha^2}{k \cdot v_\alpha} c_\alpha^a\left\{-v_\alpha \cdot v_\beta\left(q_\beta^\mu-\frac{k \cdot q_\beta}{k \cdot v_\alpha} v_\alpha^\mu\right)+k \cdot v_\alpha v_\beta^\mu-k \cdot v_\beta v_\alpha^\mu\right\}\right. \\
\left.+i f^{a b c} c_\alpha^b c_\beta^c\left\{2\left(k \cdot v_\beta\right) v_\alpha^\mu-\left(v_\alpha \cdot v_\beta\right) q_\alpha^\mu+\left(v_\alpha \cdot v_\beta\right) \frac{q_\alpha^2}{k \cdot v_\alpha} v_\alpha^\mu\right\}\right]~.
\end{gathered}
\end{equation}
Here $g$ is the gauge coupling, $k$ is the four-momentum of the emitted gluon such that $k^2=0$ on-shell and $\m_{\alpha,\beta}(k)$ is the integration measure
\begin{equation}
\mu_{\alpha, \beta}(k)=\frac{d^4q_\alpha}{\(2\pi\)^4}\frac{d^4q_\beta}{\(2\pi\)^4}\left[(2 \pi) \delta\left(v_\alpha \cdot q_\alpha\right) \frac{e^{i q_\alpha \cdot b_\alpha}}{q_\alpha^2}\right]\left[(2 \pi) \delta\left(v_\beta \cdot q_\beta\right) \frac{e^{i q_\beta \cdot b_\beta}}{q_\beta^2}\right](2 \pi)^4 \delta^4\left(k-q_\alpha-q_\beta\right)~,
\label{measure}
\end{equation}
where the timelike velocities of the massive particles are normalized as $v_\alpha^2=1$. Further, $b_\alpha$ are spacelike vectors which are for simplicity set to be purely transverse--perpendicular to the collision axis. Goldberger and Ridgway obtained the result in \cite{Goldberger:2016iau} by perturbatively solving the coupled system of classical Yang-Mills  and Wong equations\footnote{The Wong equations govern the worldline trajectories of massive particles carrying color charge degrees of freedom in a slowly varying background gauge field~\cite{Wong:1970fu}. They can be derived systematically from the one-loop worldline effective action in QCD-see for example \cite{Mueller:2019gjj} and references therein.}. This result is for general kinematics with arbitrary $v_\alpha$; as we shall see, this will be crucial for our double copy discussion. 

Before we do so, we will first show that one obtains the QCD Lipatov vertex by taking the ultrarelativistic limit of Eq.~\eqref{GRMain}. We first parametrize the four-velocity $v$ of the incoming particles moving with velocity $\beta$ (in $c=1$ units) along the $z$ direction as 
\be
v_\alpha = \gamma\(\frac{1+\eta_\alpha \beta}{\sqrt{2}},\frac{1-\eta_\alpha \beta}{\sqrt{2}},0,0\)\equiv \sqrt{2}\,\gamma\,n_\alpha ~,
\ee
where $\gamma$ is the Lorentz boost factor $\gamma = \(1-\beta^2\)^{-1/2}$ and $\eta$ is a sign ($\eta_1=-\eta_2=1$) that is positive (negative) for a particle moving in the positive (negative) $z$ direction. In the rightmost expression, we introduced the unit four-vector $n_\alpha$ which becomes a null-vector when $\beta\to 1$. (We employ lightcone coordinates $v=(v^+,v^-,v^i)$, which satisfy $v^2 = 2v^+v^- - v_i^2 = 1$.) 

Consider $\delta(q_\alpha \cdot v_\alpha)$ in Eq.~\eqref{measure}. Using $\delta(ax) = \delta(x)/|a|$, we can write it as $\delta(q_\alpha \cdot v_\a) = \delta\(q_\alpha \cdot n_\alpha\)/(\sqrt{2}\gamma)$, allowing one to extract a $1/\gamma^2$ factor from the measure. This cancels with the corresponding $\gamma$ factors obtained when replacing the four-velocities $v_\alpha$ by the unit vector $n_\alpha$ in the two curly brackets of Eq.~\eqref{GRMain}. However the first curly bracket term has a $1/m (k\cdot v_\alpha)$ prefactor, for which the ultra-relativistic limit gives $1/m\gamma\rightarrow 0$, if the mass is kept fixed. This term therefore vanishes in the Regge limit for massive particles; only the terms in the second line of Eq.~\eqref{GRMain} survive.

Performing the $q_{1,2}$ integrations in the surviving terms (which imposes constraints $q_1^-=q_2^+=0$, $k^+ = q_1^+, k^- = q_2^-$ and $\bsk = \bsq_1+\bsq_2$ that come from $\delta$-function integrations) we obtain
\begin{align}
\label{QCDLVLCEik}
    A^{\mu,a}(k) = -\frac{g^3}{k^2}\int & \frac{d^2\bsq_{2}}{\(2\pi\)^2} \frac{e^{-i \bsq_{1} \cdot \bsb_{1}}}{\bsq_{1}^2} \frac{e^{-i \bsq_{2} \cdot \bsb_{2}}}{\bsq_{2}^2} i f^{a b c} c_1^b c_2^c \(2(k\cdot n_2) n_1^\m-2(k\cdot n_1)n_2^\m -\frac{\bsq_1^2}{k\cdot n_1}n_1^\m+\frac{\bsq_2^2}{k\cdot n_2}n_2^\m-q_1^\m + q_2^\m\).
\end{align}
Writing this equation in terms of the momenta of the incoming external color charges by converting $n_\alpha$ (which are now null-vectors since we have taken the $\beta\to 1$ limit) to $p_\alpha$,
\be
\label{misl1}
n_1^\m n_2^\n = \frac{p_1^\m p_2^\n}{p_1\cdot p_2}~,\qquad \frac{n_\alpha^\m}{k\cdot n_\alpha} = \frac{p_\alpha^\m}{k\cdot p_\alpha}
\ee
and using the relations
\be
\label{misl2}
q_1^\m = \bsq_1^\m +\frac{k\cdot p_2}{p_1\cdot p_2}p_1^\m~,\qquad q_2^\m = \bsq_2^\m +\frac{k\cdot p_1}{p_1\cdot p_2}p_2^\m~,
\ee
the QCD Lipatov vertex (defined in Eq.~\eqref{LipatovVertexQCD}) is manifestly recovered in the resulting expression,
\begin{align}
\label{QCDLVLCEik1}
    A^{\mu,a}(k) = -\frac{g^3}{k^2}\int & \frac{d^2\bsq_{2}}{\(2\pi\)^2} \frac{e^{-i \bsq_{1} \cdot \bsb_{1}}}{\bsq_{1}^2} \frac{e^{-i \bsq_{2} \cdot \bsb_{2}}}{\bsq_{2}^2} i f^{a b c} \,c_1^b \,c_2^c \,\,C^\m(\bsq_1, \bsq_2)~.
\end{align}

We will now show how one obtains the amplitude for gravitational radiation in ultrarelativistic shockwave collisions using the classical double copy. Importantly, one needs to work with the general expression in Eq.~\eqref{GRMain}, impose the classical color-kinematic duality, and then take the ultrarelativistic limit of $\beta\rightarrow 1$. Working directly with Eq.~\eqref{QCDLVLCEik1} will give the wrong result -- the order of limits matter! In other words, one needs to keep the first term in Eq.~\eqref{GRMain}. 

Writing the external momentum $p_\alpha=m_\alpha v_\alpha$, which is finite but corresponds to $\gamma \gg 1$,  Eq.~\eqref{GRMain} can be expressed as 
\begin{align}
\label{QCDLVLCSubEik}
    &A^{\mu,a}(k) = -\frac{g^3}{k^2}\int  \frac{d^2\bsq_{2}}{\(2\pi\)^2} \frac{e^{-i \bsq_{1} \cdot \bsb_{1}}}{\bsq_{1}^2} \frac{e^{-i \bsq_{2} \cdot \bsb_{2}}}{\bsq_{2}^2} \bigg[i f^{a b c} c_1^b c_2^c \(2\frac{k\cdot p_2}{p_1\cdot p_2} p_1^\m-2\frac{k\cdot n_1}{p_1\cdot p_2}p_2^\m +\frac{q_1^2}{k\cdot p_1}p_1^\m-\frac{q_2^2}{k\cdot p_2}p_2^\m-q_1^\m + q_2^\m\) \\[5pt]
    & +c_1 \cdot c_2 \bigg\{ \frac{q_1^2 c_1^a}{p_1\cdot k}\(-q_2^\mu +\frac{k\cdot q_2}{k\cdot p_1} p_1^\m +\frac{k\cdot p_1}{p_1\cdot p_2}p_2^\m - \frac{k\cdot p_2}{p_1\cdot p_2} p_1^\m\)+ \frac{q_2^2 c_2^a}{p_2\cdot k}\(-q_1^\mu +\frac{k\cdot q_1}{k\cdot p_2} p_2^\m +\frac{k\cdot p_2}{p_1\cdot p_2}p_1^\m - \frac{k\cdot p_1}{p_1\cdot p_2} p_2^\m\)\bigg\}\bigg]~.\no
\end{align}
The first line in the square bracket of this expression is the universal QCD 
Lipatov vertex demonstrated previously. The second line contains non-universal\footnote{These terms are non-universal because they are sensitive to the nature of the external particles, for instance their spin.} sub-eikonal $1/p_1^+$ and $1/p_2^-$ corrections. 

The problem of identifying\footnote{
If one instead applies the usual BCJ color-kinematic duality relations to the scattering of colored scalar fields in a theory of Yang-Mills coupled to matter, one does not recover the Lipatov vertex, as discussed at length in \cite{SabioVera:2011wy, SabioVera:2012zky, Johansson:2013nsa}.}  and applying the appropriate classical BCJ-type color-kinematic prescription \cite{Bern:2008qj, Bern:2019prr} to this expression was discussed in \cite{Goldberger:2016iau}. Their replacement rules are as follows\footnote{These rules have been generalized to higher orders in perturbation theory~\cite{Shen:2018ebu}. For a review of recent developments, see \cite{Adamo:2022dcm}.}:
\begin{align}
\begin{split}
\label{DCprescription}
    &c^a_\a \to p^\m_\a  ~, \\
    &i f^{a_1 a_2 a_3} \rightarrow \Gamma^{\nu_1 \nu_2 \nu_3}\left(q_1, q_2, q_3\right)=-\frac{1}{2}\left(\eta^{\nu_1 \nu_3}\left(q_1-q_3\right)^{\nu_2}+\eta^{\nu_1 \nu_2}\left(q_2-q_1\right)^{\nu_3}+\eta^{\nu_2 \nu_3}\left(q_3-q_2\right)^{\nu_1}\right)~,  \\
    &g \to \kappa ~,
\end{split}
\end{align}
where $\kappa$ is related to Newton's constant. As explained in \cite{Goldberger:2016iau}, the motivation behind proposing the first color-kinematic replacement comes from the structural similarity between the classical equations of motion in QCD and gravity. Evidence for this replacement also comes from analyzing the double copy structure of classical fields due to boosted sources \cite{Akhoury:2013yua}, and radiation in the soft regime \cite{PV:2019uuv}. 

Applying Eq.~\eqref{DCprescription} to Eq.~\eqref{QCDLVLCSubEik} gives $A^{\mu,a}\rightarrow A^{\mu\nu}$, the gravitational wave amplitude 
\begin{align}
\label{QCDLVLCSubEikGrav}
    A^{\mu\nu}(k) = -\frac{\kappa^3}{k^2}\int & \frac{d^2\bsq_{2}}{\(2\pi\)^2} \frac{e^{-i \bsq_{1} \cdot \bsb_{1}}}{\bsq_{1}^2} \frac{e^{-i \bsq_{2} \cdot \bsb_{2}}}{\bsq_{2}^2} \bigg[-\frac12 \(-2 p_2^\n (k \cdot p_1)+2 p_1^\n (k \cdot p_2)+\frac s2(q_2-q_1)^\n\)\no\\
    &\(-\bsq_1^\m + \bsq_2^\m + \frac{k\cdot p_2}{p_1\cdot p_2} p_1^\m-\frac{k\cdot n_1}{p_1\cdot p_2}p_2^\m -\frac{\bsq_1^2}{k\cdot p_1}p_1^\m+\frac{\bsq_2^2}{k\cdot p_2}p_2^\m\) \no\\[5pt]
    & +\frac{s}{2} \bigg\{ \frac{q_1^2 p_1^\n}{p_1\cdot k}\(-\bsq_2^\mu +\frac{k\cdot q_2}{k\cdot p_1} p_1^\m - \frac{k\cdot p_2}{p_1\cdot p_2} p_1^\m\) + \frac{q_2^2 p_2^\n}{p_2\cdot k}\(-\bsq_1^\mu +\frac{k\cdot q_1}{k\cdot p_2} p_2^\m - \frac{k\cdot p_1}{p_1\cdot p_2} p_2^\m\)\bigg\}\bigg]~.
\end{align}
The first line in the square brackets is obtained from the replacement rule for the structure constants where the momenta $q_1,q_2$ are incoming and $k=q_1+q_2$ is outgoing. In writing this expression, we used the constraint $n_\alpha\cdot q_\alpha=0$ and replaced $2p_1\cdot p_2$ with the squared center-of-mass energy  $s$. The second line is simply the parenthesis in the first line of Eq.~\eqref{QCDLVLCSubEik} where we additionally used Eq.~\eqref{misl2} to express $q_\alpha^\m$ in terms of the transverse momenta alone. When written in this way, the second line is identical to the expression for the QCD Lipatov vertex in Eq.~\eqref{LipatovVertexQCD}. The last line is the same as in Eq.~\eqref{QCDLVLCSubEik} but with the replacement $c^a \to p^\m$ and the further simplifications arising (as for the second line) from using Eq.~\eqref{misl2}. 

 An important observation is that the terms in the second line of Eq.~\eqref{QCDLVLCSubEik} that were subleading in the QCD case for large $p_{1}^+$, $p_{2}^-$ now contribute at leading order. This is a direct consequence of the color-kinematic replacement $c^a\to p^\m$ and demonstrates that the eikonal approximation does not commute with the double copy procedure. To emphasize our  earlier remark, in the Regge limit, one needs to keep the appropriate sub-eikonal terms in the QCD result before performing the double copy procedure to arrive at the gravity result. 

The quickest way to see that  Eq.~\eqref{QCDLVLCSubEikGrav} indeed contains the gravitational Lipatov vertex is to contract the expression with the physical graviton polarization tensor in the lightcone gauge that is set by demanding $\epsilon_{\m\n} p_1^\m = 0$. Upon setting the momenta of the emitted graviton on-shell $k^2=0$, one finds
\begin{align}
\label{QCDLVLCSubEikGravLC}
    \epsilon_{\m\n}A^{\m\n}(k) &= \frac{\kappa^3 s}{k^2} \int \frac{d^2\bsq_{2}}{\(2\pi\)^2} \frac{e^{-i \bsq_{1} \cdot \bsb_{1}}}{\bsq_{1}^2}\frac{e^{-i \bsq_{2} \cdot \bsb_{2}}}{\bsq_{2}^2} \[\(\bsq_2^i-k^i\frac{\bsq_{2}^2}{\bsk^2}\)\(\bsq_2^j-k^j\frac{\bsq_{2}^2}{\bsk^2}\)-\frac{k^ik^j}{\bsk^4}\bsq_{1}^2\bsq_{2}^2\]\epsilon_{ij}~.
\end{align}
The expression in square brackets is precisely the lightcone gauge form of the gravitational Lipatov vertex in Eq.~\eqref{LipatovVertexGrav} determining the radiation amplitude for 
gravitational waves emerging from shockwave collisions that was computed in \cite{Raj:2023irr}. Alternatively, one can simplify the fully covariant expression given in Eq.~\eqref{QCDLVLCSubEikGrav}. Straightforward algebra gives 
\begin{align}
\label{QCDLVLCSubEikGrav1}
    A^{\mu\nu}(k) = \frac{\kappa^3s}{2\,k^2}\int & \frac{d^2\bsq_{2}}{\(2\pi\)^2} \frac{e^{-i \bsq_{1} \cdot \bsb_{1}}}{\bsq_{1}^2} \frac{e^{-i \bsq_{2} \cdot \bsb_{2}}}{\bsq_{2}^2} \frac{1}{2}\,\[C^\m C^\n - N^\m N^\n + k^\m\(\frac{p_1^\n}{p_1\cdot k}\bsq_1^2+\frac{p_2^\n}{p_2\cdot k}\bsq_2^2\)\]~.
\end{align}
The term in the bracket  that is proportional to $k^\m$ is unphysical and can be removed by a gauge transformation. When contracted with the graviton polarization tensor this term drops and therefore does not contribute to physical quantities like the intensity of gravitational wave radiation. As discussed earlier, the classical double copy procedure in general leads to results in dilaton gravity and not pure Einstein gravity. Therefore one might expect the result in Eq.~\eqref{QCDLVLCSubEikGrav1} to include a contribution from dilaton exchange. However, in a previous paper \cite{Raj:2023irr}, we showed that the expression in Eq.~\eqref{QCDLVLCSubEikGravLC} is a result in pure gravity, confirming that there is no contribution to Eq.~\eqref{QCDLVLCSubEikGrav1} from  dilaton exchange. This is of course an alternative proof to that of \cite{Goldberger_2017} who also demonstrated the decoupling of the dilaton in the high energy ultrarelativistic limit,\cite{Goldberger_2017}.

To summarize, employing the Yang-Mills + Wong equation framework for gluon radiation discussed in \cite{Goldberger:2016iau}, implementing  the classical double copy prescription, and appropriately taking the ultrarelativistic limit correctly reproduces the leading contribution to the gravitational radiation computed using Feynman diagram techniques in \cite{Lipatov:1982it,Lipatov:1982vv,Amati:1993tb} and from solutions of Einstein's equations for shockwave collisions in \cite{Raj:2023irr}. A striking feature of our result is that since the sub-eikonal terms in QCD are in general not universal, the correct universal gravitational answer appears to be dual to the particular dynamics in QCD represented by the Yang-Mills + Wong equations. A further point addressing a question raised in \cite{Luna:2017dtq} in the context of Regge asymptotics is that since the Lipatov vertex is a key piece of the $2\rightarrow N$ scattering amplitude, BCJ-type color-kinematic relations may allow one to map $2\rightarrow N$ QCD amplitudes {\it a la BFKL} to their counterparts in gravity.  

In our discussion thus far, we considered only the double copy features of  the hard emission vertices where the momentum $k$ of emitted graviton/gluon is large. For multi-Regge kinematics in QCD, this would correspond to $|\bsk|\gg\Lambda_{\rm QCD}$, where the weak coupling expansion is meaningful. Formally however one could examine the small $k$ limit of these formulas, as was done in \cite{PV:2019uuv}. It was shown there that in this limit the gluon emission vertex Eq.~\eqref{GRMain} and its gravitational counterpart (obtained by the classical double copy replacement) Eq.~\eqref{DCprescription} reduces to the well-known soft gluon and soft graviton factors respectively. The latter was first derived in  \cite{Weinberg:1965nx}. As emphasized by Lipatov in \cite{Lipatov:1982vv}, these soft factors factorize from the hard matrix elements containing the Lipatov vertex in both QCD and gravity; see also \cite{Addazi:2016ksu} for a relevant discussion. It is remarkable 
that all of this information is implicitly contained in Eq.~\eqref{GRMain}. 

A natural extension of this work is to explore the application of the classical color-kinematic double copy to derive gravitational shockwave propagators~\cite{Galley:2013eba} from the QCD shockwave propagators ~\cite{McLerran:1994vd,Ayala:1995kg,McLerran:1998nk,Balitsky:2001mr} computed in the Color Glass Condensate classical\footnote{Note that ``classical" in the CGC EFT refers specifically to modes with high occupancies as opposed to classical world line trajectories discussed extensively in the double copy literature~\cite{Kosower:2018adc}. The relation between the two frameworks has been discussed previously in \cite{delaCruz:2020bbn} and deserves further attention.} effective theory (CGC EFT) of high energy QCD~\cite{Gelis:2010nm}. As  outlined in \cite{Raj:2023irr}, a longer term goal is a quantitative approach to the CGC-Black Hole correspondence proposed in \cite{Dvali:2021ooc}.

\begin{acknowledgments}
We  thank to P. V. Athira, Leonardo de la Cruz, Nava Gaddam and Siddharth Prabhu for discussions. R.V is supported by the U.S. Department of Energy, Office of Science under contract DE-SC0012704 and within the framework of the SURGE Topical Theory Collaboration. He acknowledges partial support from an LDRD at BNL. R.V was also supported at Stony Brook by the Simons Foundation as a co-PI under Award number 994318 (Simons Collaboration on Confinement and QCD Strings). H.R is a Simons Foundation postdoctoral fellow at Stony Brook supported under Award number 994318. 
\end{acknowledgments}

\bibliography{apssamp}

\begin{thebibliography}{40}%
\makeatletter
\providecommand \@ifxundefined [1]{%
 \@ifx{#1\undefined}
}%
\providecommand \@ifnum [1]{%
 \ifnum #1\expandafter \@firstoftwo
 \else \expandafter \@secondoftwo
 \fi
}%
\providecommand \@ifx [1]{%
 \ifx #1\expandafter \@firstoftwo
 \else \expandafter \@secondoftwo
 \fi
}%
\providecommand \natexlab [1]{#1}%
\providecommand \enquote  [1]{``#1''}%
\providecommand \bibnamefont  [1]{#1}%
\providecommand \bibfnamefont [1]{#1}%
\providecommand \citenamefont [1]{#1}%
\providecommand \href@noop [0]{\@secondoftwo}%
\providecommand \href [0]{\begingroup \@sanitize@url \@href}%
\providecommand \@href[1]{\@@startlink{#1}\@@href}%
\providecommand \@@href[1]{\endgroup#1\@@endlink}%
\providecommand \@sanitize@url [0]{\catcode `\\12\catcode `\$12\catcode
  `\&12\catcode `\#12\catcode `\^12\catcode `\_12\catcode `\%12\relax}%
\providecommand \@@startlink[1]{}%
\providecommand \@@endlink[0]{}%
\providecommand \url  [0]{\begingroup\@sanitize@url \@url }%
\providecommand \@url [1]{\endgroup\@href {#1}{\urlprefix }}%
\providecommand \urlprefix  [0]{URL }%
\providecommand \Eprint [0]{\href }%
\providecommand \doibase [0]{http://dx.doi.org/}%
\providecommand \selectlanguage [0]{\@gobble}%
\providecommand \bibinfo  [0]{\@secondoftwo}%
\providecommand \bibfield  [0]{\@secondoftwo}%
\providecommand \translation [1]{[#1]}%
\providecommand \BibitemOpen [0]{}%
\providecommand \bibitemStop [0]{}%
\providecommand \bibitemNoStop [0]{.\EOS\space}%
\providecommand \EOS [0]{\spacefactor3000\relax}%
\providecommand \BibitemShut  [1]{\csname bibitem#1\endcsname}%
\let\auto@bib@innerbib\@empty
\bibitem [{\citenamefont {Raj}\ and\ \citenamefont
  {Venugopalan}(2024)}]{Raj:2023irr}%
  \BibitemOpen
  \bibfield  {author} {\bibinfo {author} {\bibfnamefont {H.}~\bibnamefont
  {Raj}}\ and\ \bibinfo {author} {\bibfnamefont {R.}~\bibnamefont
  {Venugopalan}},\ }\href {\doibase 10.1103/PhysRevD.109.044064} {\bibfield
  {journal} {\bibinfo  {journal} {Phys. Rev. D}\ }\textbf {\bibinfo {volume}
  {109}},\ \bibinfo {pages} {044064} (\bibinfo {year} {2024})},\ \Eprint
  {http://arxiv.org/abs/2311.03463} {arXiv:2311.03463 [hep-th]} \BibitemShut
  {NoStop}%
\bibitem [{\citenamefont {Lipatov}(1982{\natexlab{a}})}]{Lipatov:1982it}%
  \BibitemOpen
  \bibfield  {author} {\bibinfo {author} {\bibfnamefont {L.~N.}\ \bibnamefont
  {Lipatov}},\ }\href@noop {} {\bibfield  {journal} {\bibinfo  {journal} {Sov.
  Phys. JETP}\ }\textbf {\bibinfo {volume} {55}},\ \bibinfo {pages} {582}
  (\bibinfo {year} {1982}{\natexlab{a}})}\BibitemShut {NoStop}%
\bibitem [{\citenamefont {Lipatov}(1988)}]{Lipatov:1988ii}%
  \BibitemOpen
  \bibfield  {author} {\bibinfo {author} {\bibfnamefont {L.~N.}\ \bibnamefont
  {Lipatov}},\ }\href {\doibase 10.1016/0550-3213(88)90105-8} {\bibfield
  {journal} {\bibinfo  {journal} {Nucl. Phys. B}\ }\textbf {\bibinfo {volume}
  {307}},\ \bibinfo {pages} {705} (\bibinfo {year} {1988})}\BibitemShut
  {NoStop}%
\bibitem [{\citenamefont {Amati}\ \emph {et~al.}(1988)\citenamefont {Amati},
  \citenamefont {Ciafaloni},\ and\ \citenamefont {Veneziano}}]{Amati:1987uf}%
  \BibitemOpen
  \bibfield  {author} {\bibinfo {author} {\bibfnamefont {D.}~\bibnamefont
  {Amati}}, \bibinfo {author} {\bibfnamefont {M.}~\bibnamefont {Ciafaloni}}, \
  and\ \bibinfo {author} {\bibfnamefont {G.}~\bibnamefont {Veneziano}},\ }\href
  {\doibase 10.1142/S0217751X88000710} {\bibfield  {journal} {\bibinfo
  {journal} {Int. J. Mod. Phys. A}\ }\textbf {\bibinfo {volume} {3}},\ \bibinfo
  {pages} {1615} (\bibinfo {year} {1988})}\BibitemShut {NoStop}%
\bibitem [{\citenamefont {Lipatov}(1991)}]{Lipatov:1991nf}%
  \BibitemOpen
  \bibfield  {author} {\bibinfo {author} {\bibfnamefont {L.~N.}\ \bibnamefont
  {Lipatov}},\ }\href {\doibase 10.1016/0550-3213(91)90512-V} {\bibfield
  {journal} {\bibinfo  {journal} {Nucl. Phys. B}\ }\textbf {\bibinfo {volume}
  {365}},\ \bibinfo {pages} {614} (\bibinfo {year} {1991})}\BibitemShut
  {NoStop}%
\bibitem [{\citenamefont {Amati}\ \emph {et~al.}(1993)\citenamefont {Amati},
  \citenamefont {Ciafaloni},\ and\ \citenamefont {Veneziano}}]{Amati:1993tb}%
  \BibitemOpen
  \bibfield  {author} {\bibinfo {author} {\bibfnamefont {D.}~\bibnamefont
  {Amati}}, \bibinfo {author} {\bibfnamefont {M.}~\bibnamefont {Ciafaloni}}, \
  and\ \bibinfo {author} {\bibfnamefont {G.}~\bibnamefont {Veneziano}},\ }\href
  {\doibase 10.1016/0550-3213(93)90367-X} {\bibfield  {journal} {\bibinfo
  {journal} {Nucl. Phys. B}\ }\textbf {\bibinfo {volume} {403}},\ \bibinfo
  {pages} {707} (\bibinfo {year} {1993})}\BibitemShut {NoStop}%
\bibitem [{\citenamefont {Kuraev}\ \emph {et~al.}(1977)\citenamefont {Kuraev},
  \citenamefont {Lipatov},\ and\ \citenamefont {Fadin}}]{Kuraev:1977fs}%
  \BibitemOpen
  \bibfield  {author} {\bibinfo {author} {\bibfnamefont {E.~A.}\ \bibnamefont
  {Kuraev}}, \bibinfo {author} {\bibfnamefont {L.~N.}\ \bibnamefont {Lipatov}},
  \ and\ \bibinfo {author} {\bibfnamefont {V.~S.}\ \bibnamefont {Fadin}},\
  }\href@noop {} {\bibfield  {journal} {\bibinfo  {journal} {Sov. Phys. JETP}\
  }\textbf {\bibinfo {volume} {45}},\ \bibinfo {pages} {199} (\bibinfo {year}
  {1977})}\BibitemShut {NoStop}%
\bibitem [{\citenamefont {Balitsky}\ and\ \citenamefont
  {Lipatov}(1978)}]{Balitsky:1978ic}%
  \BibitemOpen
  \bibfield  {author} {\bibinfo {author} {\bibfnamefont {I.~I.}\ \bibnamefont
  {Balitsky}}\ and\ \bibinfo {author} {\bibfnamefont {L.~N.}\ \bibnamefont
  {Lipatov}},\ }\href@noop {} {\bibfield  {journal} {\bibinfo  {journal} {Sov.
  J. Nucl. Phys.}\ }\textbf {\bibinfo {volume} {28}},\ \bibinfo {pages} {822}
  (\bibinfo {year} {1978})}\BibitemShut {NoStop}%
\bibitem [{\citenamefont {Del~Duca}(1995)}]{DelDuca:1995hf}%
  \BibitemOpen
  \bibfield  {author} {\bibinfo {author} {\bibfnamefont {V.}~\bibnamefont
  {Del~Duca}},\ }\href@noop {} {\  (\bibinfo {year} {1995})},\ \Eprint
  {http://arxiv.org/abs/hep-ph/9503226} {arXiv:hep-ph/9503226} \BibitemShut
  {NoStop}%
\bibitem [{\citenamefont {Blaizot}\ \emph {et~al.}(2004)\citenamefont
  {Blaizot}, \citenamefont {Gelis},\ and\ \citenamefont
  {Venugopalan}}]{Blaizot:2004wu}%
  \BibitemOpen
  \bibfield  {author} {\bibinfo {author} {\bibfnamefont {J.~P.}\ \bibnamefont
  {Blaizot}}, \bibinfo {author} {\bibfnamefont {F.}~\bibnamefont {Gelis}}, \
  and\ \bibinfo {author} {\bibfnamefont {R.}~\bibnamefont {Venugopalan}},\
  }\href {\doibase 10.1016/j.nuclphysa.2004.07.005} {\bibfield  {journal}
  {\bibinfo  {journal} {Nucl. Phys. A}\ }\textbf {\bibinfo {volume} {743}},\
  \bibinfo {pages} {13} (\bibinfo {year} {2004})},\ \Eprint
  {http://arxiv.org/abs/hep-ph/0402256} {arXiv:hep-ph/0402256} \BibitemShut
  {NoStop}%
\bibitem [{\citenamefont {Gelis}\ and\ \citenamefont
  {Mehtar-Tani}(2006)}]{Gelis:2005pt}%
  \BibitemOpen
  \bibfield  {author} {\bibinfo {author} {\bibfnamefont {F.}~\bibnamefont
  {Gelis}}\ and\ \bibinfo {author} {\bibfnamefont {Y.}~\bibnamefont
  {Mehtar-Tani}},\ }\href {\doibase 10.1103/PhysRevD.73.034019} {\bibfield
  {journal} {\bibinfo  {journal} {Phys. Rev. D}\ }\textbf {\bibinfo {volume}
  {73}},\ \bibinfo {pages} {034019} (\bibinfo {year} {2006})},\ \Eprint
  {http://arxiv.org/abs/hep-ph/0512079} {arXiv:hep-ph/0512079} \BibitemShut
  {NoStop}%
\bibitem [{\citenamefont {Berges}\ \emph {et~al.}(2021)\citenamefont {Berges},
  \citenamefont {Heller}, \citenamefont {Mazeliauskas},\ and\ \citenamefont
  {Venugopalan}}]{Berges:2020fwq}%
  \BibitemOpen
  \bibfield  {author} {\bibinfo {author} {\bibfnamefont {J.}~\bibnamefont
  {Berges}}, \bibinfo {author} {\bibfnamefont {M.~P.}\ \bibnamefont {Heller}},
  \bibinfo {author} {\bibfnamefont {A.}~\bibnamefont {Mazeliauskas}}, \ and\
  \bibinfo {author} {\bibfnamefont {R.}~\bibnamefont {Venugopalan}},\ }\href
  {\doibase 10.1103/RevModPhys.93.035003} {\bibfield  {journal} {\bibinfo
  {journal} {Rev. Mod. Phys.}\ }\textbf {\bibinfo {volume} {93}},\ \bibinfo
  {pages} {035003} (\bibinfo {year} {2021})},\ \Eprint
  {http://arxiv.org/abs/2005.12299} {arXiv:2005.12299 [hep-th]} \BibitemShut
  {NoStop}%
\bibitem [{\citenamefont {Monteiro}\ \emph {et~al.}(2014)\citenamefont
  {Monteiro}, \citenamefont {O'Connell},\ and\ \citenamefont
  {White}}]{Monteiro:2014cda}%
  \BibitemOpen
  \bibfield  {author} {\bibinfo {author} {\bibfnamefont {R.}~\bibnamefont
  {Monteiro}}, \bibinfo {author} {\bibfnamefont {D.}~\bibnamefont {O'Connell}},
  \ and\ \bibinfo {author} {\bibfnamefont {C.~D.}\ \bibnamefont {White}},\
  }\href {\doibase 10.1007/JHEP12(2014)056} {\bibfield  {journal} {\bibinfo
  {journal} {JHEP}\ }\textbf {\bibinfo {volume} {12}},\ \bibinfo {pages} {056}
  (\bibinfo {year} {2014})},\ \Eprint {http://arxiv.org/abs/1410.0239}
  {arXiv:1410.0239 [hep-th]} \BibitemShut {NoStop}%
\bibitem [{\citenamefont {Goldberger}\ and\ \citenamefont
  {Ridgway}(2017{\natexlab{a}})}]{Goldberger:2016iau}%
  \BibitemOpen
  \bibfield  {author} {\bibinfo {author} {\bibfnamefont {W.~D.}\ \bibnamefont
  {Goldberger}}\ and\ \bibinfo {author} {\bibfnamefont {A.~K.}\ \bibnamefont
  {Ridgway}},\ }\href {\doibase 10.1103/PhysRevD.95.125010} {\bibfield
  {journal} {\bibinfo  {journal} {Phys. Rev. D}\ }\textbf {\bibinfo {volume}
  {95}},\ \bibinfo {pages} {125010} (\bibinfo {year} {2017}{\natexlab{a}})},\
  \Eprint {http://arxiv.org/abs/1611.03493} {arXiv:1611.03493 [hep-th]}
  \BibitemShut {NoStop}%
\bibitem [{\citenamefont {Bern}\ \emph {et~al.}(2008)\citenamefont {Bern},
  \citenamefont {Carrasco},\ and\ \citenamefont {Johansson}}]{Bern:2008qj}%
  \BibitemOpen
  \bibfield  {author} {\bibinfo {author} {\bibfnamefont {Z.}~\bibnamefont
  {Bern}}, \bibinfo {author} {\bibfnamefont {J.~J.~M.}\ \bibnamefont
  {Carrasco}}, \ and\ \bibinfo {author} {\bibfnamefont {H.}~\bibnamefont
  {Johansson}},\ }\href {\doibase 10.1103/PhysRevD.78.085011} {\bibfield
  {journal} {\bibinfo  {journal} {Phys. Rev. D}\ }\textbf {\bibinfo {volume}
  {78}},\ \bibinfo {pages} {085011} (\bibinfo {year} {2008})},\ \Eprint
  {http://arxiv.org/abs/0805.3993} {arXiv:0805.3993 [hep-ph]} \BibitemShut
  {NoStop}%
\bibitem [{\citenamefont {Bern}\ \emph {et~al.}(2019)\citenamefont {Bern},
  \citenamefont {Carrasco}, \citenamefont {Chiodaroli}, \citenamefont
  {Johansson},\ and\ \citenamefont {Roiban}}]{Bern:2019prr}%
  \BibitemOpen
  \bibfield  {author} {\bibinfo {author} {\bibfnamefont {Z.}~\bibnamefont
  {Bern}}, \bibinfo {author} {\bibfnamefont {J.~J.}\ \bibnamefont {Carrasco}},
  \bibinfo {author} {\bibfnamefont {M.}~\bibnamefont {Chiodaroli}}, \bibinfo
  {author} {\bibfnamefont {H.}~\bibnamefont {Johansson}}, \ and\ \bibinfo
  {author} {\bibfnamefont {R.}~\bibnamefont {Roiban}},\ }\href@noop {} {\
  (\bibinfo {year} {2019})},\ \Eprint {http://arxiv.org/abs/1909.01358}
  {arXiv:1909.01358 [hep-th]} \BibitemShut {NoStop}%
\bibitem [{\citenamefont {Goldberger}\ \emph {et~al.}(2018)\citenamefont
  {Goldberger}, \citenamefont {Li},\ and\ \citenamefont
  {Prabhu}}]{Goldberger:2017ogt}%
  \BibitemOpen
  \bibfield  {author} {\bibinfo {author} {\bibfnamefont {W.~D.}\ \bibnamefont
  {Goldberger}}, \bibinfo {author} {\bibfnamefont {J.}~\bibnamefont {Li}}, \
  and\ \bibinfo {author} {\bibfnamefont {S.~G.}\ \bibnamefont {Prabhu}},\
  }\href {\doibase 10.1103/PhysRevD.97.105018} {\bibfield  {journal} {\bibinfo
  {journal} {Phys. Rev. D}\ }\textbf {\bibinfo {volume} {97}},\ \bibinfo
  {pages} {105018} (\bibinfo {year} {2018})},\ \Eprint
  {http://arxiv.org/abs/1712.09250} {arXiv:1712.09250 [hep-th]} \BibitemShut
  {NoStop}%
\bibitem [{\citenamefont {Wong}(1970)}]{Wong:1970fu}%
  \BibitemOpen
  \bibfield  {author} {\bibinfo {author} {\bibfnamefont {S.~K.}\ \bibnamefont
  {Wong}},\ }\href {\doibase 10.1007/BF02892134} {\bibfield  {journal}
  {\bibinfo  {journal} {Nuovo Cim. A}\ }\textbf {\bibinfo {volume} {65}},\
  \bibinfo {pages} {689} (\bibinfo {year} {1970})}\BibitemShut {NoStop}%
\bibitem [{\citenamefont {Mueller}\ and\ \citenamefont
  {Venugopalan}(2019)}]{Mueller:2019gjj}%
  \BibitemOpen
  \bibfield  {author} {\bibinfo {author} {\bibfnamefont {N.}~\bibnamefont
  {Mueller}}\ and\ \bibinfo {author} {\bibfnamefont {R.}~\bibnamefont
  {Venugopalan}},\ }\href {\doibase 10.1103/PhysRevD.99.056003} {\bibfield
  {journal} {\bibinfo  {journal} {Phys. Rev. D}\ }\textbf {\bibinfo {volume}
  {99}},\ \bibinfo {pages} {056003} (\bibinfo {year} {2019})},\ \Eprint
  {http://arxiv.org/abs/1901.10492} {arXiv:1901.10492 [hep-th]} \BibitemShut
  {NoStop}%
\bibitem [{\citenamefont {Sabio~Vera}\ \emph {et~al.}(2012)\citenamefont
  {Sabio~Vera}, \citenamefont {Serna~Campillo},\ and\ \citenamefont
  {Vazquez-Mozo}}]{SabioVera:2011wy}%
  \BibitemOpen
  \bibfield  {author} {\bibinfo {author} {\bibfnamefont {A.}~\bibnamefont
  {Sabio~Vera}}, \bibinfo {author} {\bibfnamefont {E.}~\bibnamefont
  {Serna~Campillo}}, \ and\ \bibinfo {author} {\bibfnamefont {M.~A.}\
  \bibnamefont {Vazquez-Mozo}},\ }\href {\doibase 10.1007/JHEP03(2012)005}
  {\bibfield  {journal} {\bibinfo  {journal} {JHEP}\ }\textbf {\bibinfo
  {volume} {03}},\ \bibinfo {pages} {005} (\bibinfo {year} {2012})},\ \Eprint
  {http://arxiv.org/abs/1112.4494} {arXiv:1112.4494 [hep-th]} \BibitemShut
  {NoStop}%
\bibitem [{\citenamefont {Sabio~Vera}\ \emph {et~al.}(2013)\citenamefont
  {Sabio~Vera}, \citenamefont {Serna~Campillo},\ and\ \citenamefont
  {Vazquez-Mozo}}]{SabioVera:2012zky}%
  \BibitemOpen
  \bibfield  {author} {\bibinfo {author} {\bibfnamefont {A.}~\bibnamefont
  {Sabio~Vera}}, \bibinfo {author} {\bibfnamefont {E.}~\bibnamefont
  {Serna~Campillo}}, \ and\ \bibinfo {author} {\bibfnamefont {M.~A.}\
  \bibnamefont {Vazquez-Mozo}},\ }\href {\doibase 10.1007/JHEP04(2013)086}
  {\bibfield  {journal} {\bibinfo  {journal} {JHEP}\ }\textbf {\bibinfo
  {volume} {04}},\ \bibinfo {pages} {086} (\bibinfo {year} {2013})},\ \Eprint
  {http://arxiv.org/abs/1212.5103} {arXiv:1212.5103 [hep-th]} \BibitemShut
  {NoStop}%
\bibitem [{\citenamefont {Johansson}\ \emph {et~al.}(2013)\citenamefont
  {Johansson}, \citenamefont {Sabio~Vera}, \citenamefont {Serna~Campillo},\
  and\ \citenamefont {V\'azquez-Mozo}}]{Johansson:2013nsa}%
  \BibitemOpen
  \bibfield  {author} {\bibinfo {author} {\bibfnamefont {H.}~\bibnamefont
  {Johansson}}, \bibinfo {author} {\bibfnamefont {A.}~\bibnamefont
  {Sabio~Vera}}, \bibinfo {author} {\bibfnamefont {E.}~\bibnamefont
  {Serna~Campillo}}, \ and\ \bibinfo {author} {\bibfnamefont {M.~A.}\
  \bibnamefont {V\'azquez-Mozo}},\ }\href {\doibase 10.1007/JHEP10(2013)215}
  {\bibfield  {journal} {\bibinfo  {journal} {JHEP}\ }\textbf {\bibinfo
  {volume} {10}},\ \bibinfo {pages} {215} (\bibinfo {year} {2013})},\ \Eprint
  {http://arxiv.org/abs/1307.3106} {arXiv:1307.3106 [hep-th]} \BibitemShut
  {NoStop}%
\bibitem [{\citenamefont {Shen}(2018)}]{Shen:2018ebu}%
  \BibitemOpen
  \bibfield  {author} {\bibinfo {author} {\bibfnamefont {C.-H.}\ \bibnamefont
  {Shen}},\ }\href {\doibase 10.1007/JHEP11(2018)162} {\bibfield  {journal}
  {\bibinfo  {journal} {JHEP}\ }\textbf {\bibinfo {volume} {11}},\ \bibinfo
  {pages} {162} (\bibinfo {year} {2018})},\ \Eprint
  {http://arxiv.org/abs/1806.07388} {arXiv:1806.07388 [hep-th]} \BibitemShut
  {NoStop}%
\bibitem [{\citenamefont {Adamo}\ \emph {et~al.}(2022)\citenamefont {Adamo},
  \citenamefont {Carrasco}, \citenamefont {Carrillo-Gonz\'alez}, \citenamefont
  {Chiodaroli}, \citenamefont {Elvang}, \citenamefont {Johansson},
  \citenamefont {O'Connell}, \citenamefont {Roiban},\ and\ \citenamefont
  {Schlotterer}}]{Adamo:2022dcm}%
  \BibitemOpen
  \bibfield  {author} {\bibinfo {author} {\bibfnamefont {T.}~\bibnamefont
  {Adamo}}, \bibinfo {author} {\bibfnamefont {J.~J.~M.}\ \bibnamefont
  {Carrasco}}, \bibinfo {author} {\bibfnamefont {M.}~\bibnamefont
  {Carrillo-Gonz\'alez}}, \bibinfo {author} {\bibfnamefont {M.}~\bibnamefont
  {Chiodaroli}}, \bibinfo {author} {\bibfnamefont {H.}~\bibnamefont {Elvang}},
  \bibinfo {author} {\bibfnamefont {H.}~\bibnamefont {Johansson}}, \bibinfo
  {author} {\bibfnamefont {D.}~\bibnamefont {O'Connell}}, \bibinfo {author}
  {\bibfnamefont {R.}~\bibnamefont {Roiban}}, \ and\ \bibinfo {author}
  {\bibfnamefont {O.}~\bibnamefont {Schlotterer}},\ }in\ \href@noop {} {\emph
  {\bibinfo {booktitle} {{Snowmass 2021}}}}\ (\bibinfo {year} {2022})\ \Eprint
  {http://arxiv.org/abs/2204.06547} {arXiv:2204.06547 [hep-th]} \BibitemShut
  {NoStop}%
\bibitem [{\citenamefont {Akhoury}\ \emph {et~al.}(2021)\citenamefont
  {Akhoury}, \citenamefont {Saotome},\ and\ \citenamefont
  {Sterman}}]{Akhoury:2013yua}%
  \BibitemOpen
  \bibfield  {author} {\bibinfo {author} {\bibfnamefont {R.}~\bibnamefont
  {Akhoury}}, \bibinfo {author} {\bibfnamefont {R.}~\bibnamefont {Saotome}}, \
  and\ \bibinfo {author} {\bibfnamefont {G.}~\bibnamefont {Sterman}},\ }\href
  {\doibase 10.1103/PhysRevD.103.064036} {\bibfield  {journal} {\bibinfo
  {journal} {Phys. Rev. D}\ }\textbf {\bibinfo {volume} {103}},\ \bibinfo
  {pages} {064036} (\bibinfo {year} {2021})},\ \Eprint
  {http://arxiv.org/abs/1308.5204} {arXiv:1308.5204 [hep-th]} \BibitemShut
  {NoStop}%
\bibitem [{\citenamefont {Athira}\ and\ \citenamefont
  {Manu}(2020)}]{PV:2019uuv}%
  \BibitemOpen
  \bibfield  {author} {\bibinfo {author} {\bibfnamefont {P.~V.}\ \bibnamefont
  {Athira}}\ and\ \bibinfo {author} {\bibfnamefont {A.}~\bibnamefont {Manu}},\
  }\href {\doibase 10.1103/PhysRevD.101.046014} {\bibfield  {journal} {\bibinfo
   {journal} {Phys. Rev. D}\ }\textbf {\bibinfo {volume} {101}},\ \bibinfo
  {pages} {046014} (\bibinfo {year} {2020})},\ \Eprint
  {http://arxiv.org/abs/1907.10021} {arXiv:1907.10021 [hep-th]} \BibitemShut
  {NoStop}%
\bibitem [{\citenamefont {Goldberger}\ and\ \citenamefont
  {Ridgway}(2017{\natexlab{b}})}]{Goldberger_2017}%
  \BibitemOpen
  \bibfield  {author} {\bibinfo {author} {\bibfnamefont {W.~D.}\ \bibnamefont
  {Goldberger}}\ and\ \bibinfo {author} {\bibfnamefont {A.~K.}\ \bibnamefont
  {Ridgway}},\ }\href {\doibase 10.1103/physrevd.95.125010} {\bibfield
  {journal} {\bibinfo  {journal} {Physical Review D}\ }\textbf {\bibinfo
  {volume} {95}} (\bibinfo {year} {2017}{\natexlab{b}}),\
  10.1103/physrevd.95.125010}\BibitemShut {NoStop}%
\bibitem [{\citenamefont {Lipatov}(1982{\natexlab{b}})}]{Lipatov:1982vv}%
  \BibitemOpen
  \bibfield  {author} {\bibinfo {author} {\bibfnamefont {L.~N.}\ \bibnamefont
  {Lipatov}},\ }\href {\doibase 10.1016/0370-2693(82)90156-3} {\bibfield
  {journal} {\bibinfo  {journal} {Phys. Lett. B}\ }\textbf {\bibinfo {volume}
  {116}},\ \bibinfo {pages} {411} (\bibinfo {year}
  {1982}{\natexlab{b}})}\BibitemShut {NoStop}%
\bibitem [{\citenamefont {Luna}\ \emph {et~al.}(2018)\citenamefont {Luna},
  \citenamefont {Nicholson}, \citenamefont {O'Connell},\ and\ \citenamefont
  {White}}]{Luna:2017dtq}%
  \BibitemOpen
  \bibfield  {author} {\bibinfo {author} {\bibfnamefont {A.}~\bibnamefont
  {Luna}}, \bibinfo {author} {\bibfnamefont {I.}~\bibnamefont {Nicholson}},
  \bibinfo {author} {\bibfnamefont {D.}~\bibnamefont {O'Connell}}, \ and\
  \bibinfo {author} {\bibfnamefont {C.~D.}\ \bibnamefont {White}},\ }\href
  {\doibase 10.1007/JHEP03(2018)044} {\bibfield  {journal} {\bibinfo  {journal}
  {JHEP}\ }\textbf {\bibinfo {volume} {03}},\ \bibinfo {pages} {044} (\bibinfo
  {year} {2018})},\ \Eprint {http://arxiv.org/abs/1711.03901} {arXiv:1711.03901
  [hep-th]} \BibitemShut {NoStop}%
\bibitem [{\citenamefont {Weinberg}(1965)}]{Weinberg:1965nx}%
  \BibitemOpen
  \bibfield  {author} {\bibinfo {author} {\bibfnamefont {S.}~\bibnamefont
  {Weinberg}},\ }\href {\doibase 10.1103/PhysRev.140.B516} {\bibfield
  {journal} {\bibinfo  {journal} {Phys. Rev.}\ }\textbf {\bibinfo {volume}
  {140}},\ \bibinfo {pages} {B516} (\bibinfo {year} {1965})}\BibitemShut
  {NoStop}%
\bibitem [{\citenamefont {Addazi}\ \emph {et~al.}(2017)\citenamefont {Addazi},
  \citenamefont {Bianchi},\ and\ \citenamefont {Veneziano}}]{Addazi:2016ksu}%
  \BibitemOpen
  \bibfield  {author} {\bibinfo {author} {\bibfnamefont {A.}~\bibnamefont
  {Addazi}}, \bibinfo {author} {\bibfnamefont {M.}~\bibnamefont {Bianchi}}, \
  and\ \bibinfo {author} {\bibfnamefont {G.}~\bibnamefont {Veneziano}},\ }\href
  {\doibase 10.1007/JHEP02(2017)111} {\bibfield  {journal} {\bibinfo  {journal}
  {JHEP}\ }\textbf {\bibinfo {volume} {02}},\ \bibinfo {pages} {111} (\bibinfo
  {year} {2017})},\ \Eprint {http://arxiv.org/abs/1611.03643} {arXiv:1611.03643
  [hep-th]} \BibitemShut {NoStop}%
\bibitem [{\citenamefont {Galley}\ and\ \citenamefont
  {Porto}(2013)}]{Galley:2013eba}%
  \BibitemOpen
  \bibfield  {author} {\bibinfo {author} {\bibfnamefont {C.~R.}\ \bibnamefont
  {Galley}}\ and\ \bibinfo {author} {\bibfnamefont {R.~A.}\ \bibnamefont
  {Porto}},\ }\href {\doibase 10.1007/JHEP11(2013)096} {\bibfield  {journal}
  {\bibinfo  {journal} {JHEP}\ }\textbf {\bibinfo {volume} {11}},\ \bibinfo
  {pages} {096} (\bibinfo {year} {2013})},\ \Eprint
  {http://arxiv.org/abs/1302.4486} {arXiv:1302.4486 [gr-qc]} \BibitemShut
  {NoStop}%
\bibitem [{\citenamefont {McLerran}\ and\ \citenamefont
  {Venugopalan}(1994)}]{McLerran:1994vd}%
  \BibitemOpen
  \bibfield  {author} {\bibinfo {author} {\bibfnamefont {L.~D.}\ \bibnamefont
  {McLerran}}\ and\ \bibinfo {author} {\bibfnamefont {R.}~\bibnamefont
  {Venugopalan}},\ }\href {\doibase 10.1103/PhysRevD.50.2225} {\bibfield
  {journal} {\bibinfo  {journal} {Phys. Rev. D}\ }\textbf {\bibinfo {volume}
  {50}},\ \bibinfo {pages} {2225} (\bibinfo {year} {1994})},\ \Eprint
  {http://arxiv.org/abs/hep-ph/9402335} {arXiv:hep-ph/9402335} \BibitemShut
  {NoStop}%
\bibitem [{\citenamefont {Ayala}\ \emph {et~al.}(1995)\citenamefont {Ayala},
  \citenamefont {Jalilian-Marian}, \citenamefont {McLerran},\ and\
  \citenamefont {Venugopalan}}]{Ayala:1995kg}%
  \BibitemOpen
  \bibfield  {author} {\bibinfo {author} {\bibfnamefont {A.}~\bibnamefont
  {Ayala}}, \bibinfo {author} {\bibfnamefont {J.}~\bibnamefont
  {Jalilian-Marian}}, \bibinfo {author} {\bibfnamefont {L.~D.}\ \bibnamefont
  {McLerran}}, \ and\ \bibinfo {author} {\bibfnamefont {R.}~\bibnamefont
  {Venugopalan}},\ }\href {\doibase 10.1103/PhysRevD.52.2935} {\bibfield
  {journal} {\bibinfo  {journal} {Phys. Rev. D}\ }\textbf {\bibinfo {volume}
  {52}},\ \bibinfo {pages} {2935} (\bibinfo {year} {1995})},\ \Eprint
  {http://arxiv.org/abs/hep-ph/9501324} {arXiv:hep-ph/9501324} \BibitemShut
  {NoStop}%
\bibitem [{\citenamefont {McLerran}\ and\ \citenamefont
  {Venugopalan}(1999)}]{McLerran:1998nk}%
  \BibitemOpen
  \bibfield  {author} {\bibinfo {author} {\bibfnamefont {L.~D.}\ \bibnamefont
  {McLerran}}\ and\ \bibinfo {author} {\bibfnamefont {R.}~\bibnamefont
  {Venugopalan}},\ }\href {\doibase 10.1103/PhysRevD.59.094002} {\bibfield
  {journal} {\bibinfo  {journal} {Phys. Rev. D}\ }\textbf {\bibinfo {volume}
  {59}},\ \bibinfo {pages} {094002} (\bibinfo {year} {1999})},\ \Eprint
  {http://arxiv.org/abs/hep-ph/9809427} {arXiv:hep-ph/9809427} \BibitemShut
  {NoStop}%
\bibitem [{\citenamefont {Balitsky}\ and\ \citenamefont
  {Belitsky}(2002)}]{Balitsky:2001mr}%
  \BibitemOpen
  \bibfield  {author} {\bibinfo {author} {\bibfnamefont {I.~I.}\ \bibnamefont
  {Balitsky}}\ and\ \bibinfo {author} {\bibfnamefont {A.~V.}\ \bibnamefont
  {Belitsky}},\ }\href {\doibase 10.1016/S0550-3213(02)00149-9} {\bibfield
  {journal} {\bibinfo  {journal} {Nucl. Phys. B}\ }\textbf {\bibinfo {volume}
  {629}},\ \bibinfo {pages} {290} (\bibinfo {year} {2002})},\ \Eprint
  {http://arxiv.org/abs/hep-ph/0110158} {arXiv:hep-ph/0110158} \BibitemShut
  {NoStop}%
\bibitem [{\citenamefont {Kosower}\ \emph {et~al.}(2019)\citenamefont
  {Kosower}, \citenamefont {Maybee},\ and\ \citenamefont
  {O'Connell}}]{Kosower:2018adc}%
  \BibitemOpen
  \bibfield  {author} {\bibinfo {author} {\bibfnamefont {D.~A.}\ \bibnamefont
  {Kosower}}, \bibinfo {author} {\bibfnamefont {B.}~\bibnamefont {Maybee}}, \
  and\ \bibinfo {author} {\bibfnamefont {D.}~\bibnamefont {O'Connell}},\ }\href
  {\doibase 10.1007/JHEP02(2019)137} {\bibfield  {journal} {\bibinfo  {journal}
  {JHEP}\ }\textbf {\bibinfo {volume} {02}},\ \bibinfo {pages} {137} (\bibinfo
  {year} {2019})},\ \Eprint {http://arxiv.org/abs/1811.10950} {arXiv:1811.10950
  [hep-th]} \BibitemShut {NoStop}%
\bibitem [{\citenamefont {de~la Cruz}\ \emph {et~al.}(2020)\citenamefont {de~la
  Cruz}, \citenamefont {Maybee}, \citenamefont {O'Connell},\ and\ \citenamefont
  {Ross}}]{delaCruz:2020bbn}%
  \BibitemOpen
  \bibfield  {author} {\bibinfo {author} {\bibfnamefont {L.}~\bibnamefont
  {de~la Cruz}}, \bibinfo {author} {\bibfnamefont {B.}~\bibnamefont {Maybee}},
  \bibinfo {author} {\bibfnamefont {D.}~\bibnamefont {O'Connell}}, \ and\
  \bibinfo {author} {\bibfnamefont {A.}~\bibnamefont {Ross}},\ }\href {\doibase
  10.1007/JHEP12(2020)076} {\bibfield  {journal} {\bibinfo  {journal} {JHEP}\
  }\textbf {\bibinfo {volume} {12}},\ \bibinfo {pages} {076} (\bibinfo {year}
  {2020})},\ \Eprint {http://arxiv.org/abs/2009.03842} {arXiv:2009.03842
  [hep-th]} \BibitemShut {NoStop}%
\bibitem [{\citenamefont {Gelis}\ \emph {et~al.}(2010)\citenamefont {Gelis},
  \citenamefont {Iancu}, \citenamefont {Jalilian-Marian},\ and\ \citenamefont
  {Venugopalan}}]{Gelis:2010nm}%
  \BibitemOpen
  \bibfield  {author} {\bibinfo {author} {\bibfnamefont {F.}~\bibnamefont
  {Gelis}}, \bibinfo {author} {\bibfnamefont {E.}~\bibnamefont {Iancu}},
  \bibinfo {author} {\bibfnamefont {J.}~\bibnamefont {Jalilian-Marian}}, \ and\
  \bibinfo {author} {\bibfnamefont {R.}~\bibnamefont {Venugopalan}},\ }\href
  {\doibase 10.1146/annurev.nucl.010909.083629} {\bibfield  {journal} {\bibinfo
   {journal} {Ann. Rev. Nucl. Part. Sci.}\ }\textbf {\bibinfo {volume} {60}},\
  \bibinfo {pages} {463} (\bibinfo {year} {2010})},\ \Eprint
  {http://arxiv.org/abs/1002.0333} {arXiv:1002.0333 [hep-ph]} \BibitemShut
  {NoStop}%
\bibitem [{\citenamefont {Dvali}\ and\ \citenamefont
  {Venugopalan}(2022)}]{Dvali:2021ooc}%
  \BibitemOpen
  \bibfield  {author} {\bibinfo {author} {\bibfnamefont {G.}~\bibnamefont
  {Dvali}}\ and\ \bibinfo {author} {\bibfnamefont {R.}~\bibnamefont
  {Venugopalan}},\ }\href {\doibase 10.1103/PhysRevD.105.056026} {\bibfield
  {journal} {\bibinfo  {journal} {Phys. Rev. D}\ }\textbf {\bibinfo {volume}
  {105}},\ \bibinfo {pages} {056026} (\bibinfo {year} {2022})},\ \Eprint
  {http://arxiv.org/abs/2106.11989} {arXiv:2106.11989 [hep-th]} \BibitemShut
  {NoStop}%
\end{thebibliography}%

\end{document}